\documentclass[manuscript]{aastex631}
\shorttitle{Multiple populations in low mass globular clusters: Eridanus}
\shortauthors{Wang et al.}

\graphicspath{{./}{figures/}}
\usepackage{graphicx}
\usepackage{subfigure}
\usepackage{txfonts}
\usepackage{color}
\usepackage{ulem}
\usepackage[figuresright]{rotating}
\begin{document}

\title{Multiple Populations in Low-mass Globular Clusters: Eridanus}

\author{Yue Wang}
\affiliation{Key Laboratory of Optical Astronomy, National Astronomical Observatories, Chinese Academy of Sciences, Beijing 100101, Peoples Republic of China}

\author{Baitian Tang}
\author{Chengyuan Li}
\affiliation{School of Physics and Astronomy, Sun Yat-sen University, Zhuhai 519082, Peoples Republic of China; tangbt@mail.sysu.edu.cn}
\affiliation{CSST Science Center for the Guangdong-Hong Kong-Macau Greater Bay Area, Zhuhai, 519082, China}

\author{Holger Baumgardt}
\affiliation{School of Mathematics and Physics, The University of Queensland, St. Lucia, QLD 4072, Australia}

\author{Ricardo R. Mu\~noz}
\affiliation{Departamento de Astronomia, Universidad de Chile, Camino del Observatorio 1515, Las Condes, Santiago, Chile}

\author{Jos\'e G. Fern\'andez-Trincado}
\affiliation{Instituto de Astronom\'ia, Universidad Cat\'olica del Norte, Av. Angamos 0610, Antofagasta, Chile}

\author{Doug Geisler}
\affiliation{Departamento de Astronomia, Casilla 160-C, Universidad de Concepcion, Chile}
\affiliation{Instituto de Investigacion Multidisciplinario en Ciencia y Tecnologia, Universidad de La Serena. Avenida Raul Bitran S/N, La Serena, Chile}
\affiliation{Departamento de Astronomia, Facultad de Ciencias, Universidad de La Serena. Av. Juan Cisternas 1200, La Serena, Chile}

\author{Yuanqing Fang}
\affiliation{School of Physics and Astronomy, Sun Yat-sen University, Zhuhai 519082, Peoples Republic of China}

\begin{abstract}
Multiple populations (MPs), characterized by variations in light elemental abundances, have been found in stellar clusters in the Milky Way, Magellanic Clouds, as well as several other dwarf galaxies.
Based on a large amount of observations, mass has been suggested to be a key parameter affecting the presence and appearance of MPs in stellar clusters.
To further investigate the existence of MPs in low-mass clusters and explore the mass threshold for MP formation, we carried out a project studying the stellar population composition in several low-mass Galactic globular clusters.
Here we present our study on the cluster Eridanus. With blue-UV low-resolution spectra obtained with the OSIRIS/Multi-object spectrograph on the Gran Telescopio Canarias, we computed the spectral indices of CH and CN for the sample giant stars, and derived their carbon and nitrogen abundances using model spectra. A significant dispersion in the initial surface abundance of nitrogen was found in the sample, indicating the existence of MPs in Eridanus.
Inspecting the age-initial mass distribution of in-situ clusters with MPs, we find a slight trend that initial mass increases with increasing age, and the lowest initial mass of $\log M_{initial} \sim$ 4.98 and 5.26 are found at the young and old end, respectively, which might provide a rough reference for the mass threshold for clusters to form MPs. However, more observations of clusters with low initial masses are still necessary before any firm conclusion can be drawn.
\end{abstract}

\section{Introduction}
\label{sec:intro}

Chemical abundance variations in light elements of globular clusters (GCs) were discovered half a century ago \citep{Osborn1971}.
Besides ``normal'' stars with chemical abundances similar to field stars, the stars showing chemical anomalies, e.g., deficiency in carbon, oxygen, and magnesium; enrichment in helium, nitrogen, sodium, and aluminium, were discovered to commonly exist in GCs as well in series of observations, which are termed as ``multiple populations'' (MPs) \citep[e.g.,][and references there in]{Gratton2004,Pancino2010,Carretta2010,Pancino2017,Masseron2019,Meszaros2020}.
In addition to direct abundance analysis using spectra, photometry is also an effective way to study MPs, especially for faint clusters where high signal to noise ratio (SNR) spectra are unreachable. MP studies through photometry mainly use nitrogen-related molecular absorption bands in the UV-blue portion of the spectrum (e.g., NH, CN) \citep[e.g.,][]{Piotto2012,Piotto2015,Milone2017}.

Although the existence of MPs is currently well accepted as a general feature for GCs, theoretical explanations for its origin have not reached a consensus \citep[see][and references therein]{Bastian2018}.
Under the hypothesis of self-enrichment, which has been widely recognized, the ashes of high-temperature hydrogen burning were ejected from massive primordial stars and mixed with interstellar medium. Then enriched stars that formed out of the mixture should show the enriched chemical pattern/chemical anomalies.
The nature of the polluters has been debated for almost two decades,
e.g., massive AGB stars \citep{DErcole2010,Ventura2011}, fast-rotating massive stars \citep{Decressin2007a,Decressin2007b}, supermassive stars \citep{Denissenkov2014,Denissenkov2015}, massive interacting binaries \citep{deMink2009} and stellar mergers \citep{Wang2020}.
However, none of the popular scenarios can reproduce the complex observed phenomena \citep{Renzini2015,Bastian2018}.

After inspection of a large number of GCs, cluster mass, metallicity and age are found to be the key parameters affecting the presence of MPs \citep{Carretta2010,Bastian2018}.   To provide threshold  parameters that can discriminate between and improve existing formation models, it is vital to
investigate the detailed manifestation of MPs at critical conditions.
For example, at the metal-rich end, NGC\,6553 was found to be the most metal-rich GC with MPs, suggesting a potential metallicity ([Fe/H]) upper limit of -0.15 \citep{Tang2017}. In term of cluster age, no clusters with ages younger than $\sim$2\,Gyr have been found to host MPs so far, thus $\sim$2\,Gyr seems to be a lower limit of MP formation \citep[][and references therein]{Bastian2018}.

Cluster mass has been suggested to be a universal parameter that decides the complexity of MPs, e.g., more massive clusters have larger abundance variations and fractions of enriched populations \citep{Bastian2018,Milone2020}; more massive Type II GCs show larger iron spread \citep{Milone2022}. Exploring the critical cluster mass for MP formation will certainly enlighten the current MP scenarios.
The possible mass boundary ($\sim 5 \times 10^4 M_{\odot}$) proposed in \cite{Li2019b} seems to well separate the intermediate-age GCs (2-10 Gyr) with and without MPs. But the situation is more complicated for older GCs.
In this respect, NGC\,6535 (mass\,=\,2.2\,$\times$\,$10^{4}\,M_{\odot}$) was suspected to be the lowest-mass GC to harbor MPs based on high-resolution spectroscopy \citep{Bragaglia2017}, and moreover, MPs were found to probably exist in the lower-mass GC ESO452-SC11 (mass\,$\sim$\,(6.8\,$\pm$\,3.4)$\times\,10^{3}\,M_{\odot}$) with low-resolution spectra \citep{Simpson2017}. Palomar\,13 (Pal\,13) was recently found to show MPs based on low-resolution spectra \citep{Tang2021}, and it could probably be the lowest-mass GC (mass\,$\sim$\, $10^{3.5}\,M_{\odot}$, our most updated estimation) with MPs.

Under the hypothesis of self-enrichment, a minimum mass is indeed required by a cluster to retain ejecta from an initial generation and allow the formation of a subsequent generation, depending of course on the ejecta velocity.
For example, from hydrodynamical simulations, \citet{Vesperini2010} and \citet{Bekki2011} both found a lower mass limit, $\sim10^{4.8}-10^{5}\,M_{\odot}$ and $\sim10^{6}\,M_{\odot}$, respectively, for stellar clusters to retain enough ejecta of primordial generation stars and form enriched generation stars.
At the same time, the mass threshold has also been suggested by observations.
Based on the Na-O anticorrelations, \citet{Carretta2010} showed that a minimum present-day mass of a few $10^{4}\,M_{\odot}$ was required for stellar clusters to present MPs.
\citet{Conroy2011} found that in the Large Magellanic Cloud (LMC) only clusters more massive than about $10^{4}\,M_{\odot}$ showed MPs, and this critical mass was also consistent with their model prediction of the mass at which ram pressure was sufficient to remove the gas within the cluster.
Moreover, \citet{Milone2020} suggested a threshold of $\sim1.5\times10^{5}\,M_{\odot}$ in initial mass for Galactic GCs to form MPs, while clusters in the Magellanic Clouds seemed not to follow this limit.

Although several papers have proposed possible mass thresholds for stellar clusters to form MPs based on observations as mentioned above, no final conclusion can be derived until we have a comprehensive knowledge about the stellar population composition of low-mass clusters. However, this is a challenging work since low-mass clusters have few bright giant stars and are often located at large distances from the Sun.
To take the study one step further and have a better understanding of the existence of MPs at the low-mass end, we have observed several low-mass Galactic GCs. Since they are remote objects, low-resolution spectra of their member stars were obtained to ensure enough signal-to-noise ratio (SNR) to derive the abundances of relevant species. As a subsequent work of our last study of MPs in  Pal\,13 \citep{Tang2021}, here we present the first study of the stellar population composition in the GC Eridanus, which has a present-day mass of $1.16\times 10^{4}\,M_{\odot}$ \citep{Baumgardt2019}.

Eridanus was first discovered by \citet{Cesarsky1977}.
Having a metallicity of about [Fe/H]\,=\,-1.4 and distance of about 90\,kpc \citep[e.g.,][2010 edition; see Sect.  \ref{subsec:basicInfo} for details]{Harris1996}, Eridanus is one of the most metal-rich clusters in the outer halo. With an age of 10.5\,Gyr, it is also younger compared with inner-halo GCs at similar metallicities \citep{Beccari2012}. Hence, Eridanus is expected to have an external origin.  Although being an interesting object for studies, it has not been explored spectroscopically, nor does its MP properties, due to its large Galactocentric distance.
In this paper, we analyze the low-resolution spectra %(R$\sim2500$)
of Eridanus member stars to estimate their carbon and nitrogen abundances, searching for possible abundance variations that indicate MPs.
In Section \ref{sec:obs-red}, we outline the observations and spectral reduction briefly. In Section \ref{sec:anal-res}, we analyze the spectra of member stars carefully and show our major results. The discussion and summary are given in Sections \ref{sec:disc} and \ref{sec:sum}, respectively.

\section{Observation and Data Reduction}
\label{sec:obs-red}

\subsection{Observation}

 The observations were carried out using the OSIRIS/Multi-object spectrograph (MOS) \footnote{http://www.gtc.iac.es/instruments/osiris/osiris.php} mounted on the Gran Telescopio Canarias (GTC), Observatorio del Roque de los Muchachos, under the program GTC2-18BCNT in December of 2018.
The sample stars for observation were initially selected from the Gaia DR1 catalogue \citep{GaiaDR1} by coordinates within the region of Eridanus GC.
Given that slitlet observation forbids two targets from overlapping in the dispersion direction and brighter targets generate higher SNR spectra within a given time, 19 member stars were finally observed (Section \ref{sec:anal-res}).  The observation consists of five observing blocks (OBs) with a total exposure time of 4.88 hours (See Table \ref{tab:obs}). Using R2500U as the dispersion element, the obtained spectra cover the wavelength range of 3500$-$4600\,\AA~ with nominal spectral resolution of $\sim$2500.

\begin{table}[thp]
 \caption{Observation details.} \label{tab:obs}
 \centering
 \begin{tabular}{c c c c c}
 \hline
 \hline
observing block  &  date  & exposure time (s) &  airmass & seeing ('')    \\
 \hline
OB02  &  2018-12-08 &  1800$\times$2  &  1.70  &   1.2   \\
OB06  &  2018-12-08 &  1800$\times$2  &  1.55  &   1.1   \\
OB07  &  2018-12-08 &  1580$\times$2  &  1.58  &   1.0   \\
OB09  &  2018-12-10 &  1800$\times$2  &  1.55  &   1.2   \\
OB10  &  2018-12-10 &  1800$\times$2  &  1.62  &   1.2   \\
 \hline
 \end{tabular}
\end{table}

\subsection{Data Reduction}

\begin{figure}[ht]
   \centering
     \includegraphics[width=0.95\textwidth]{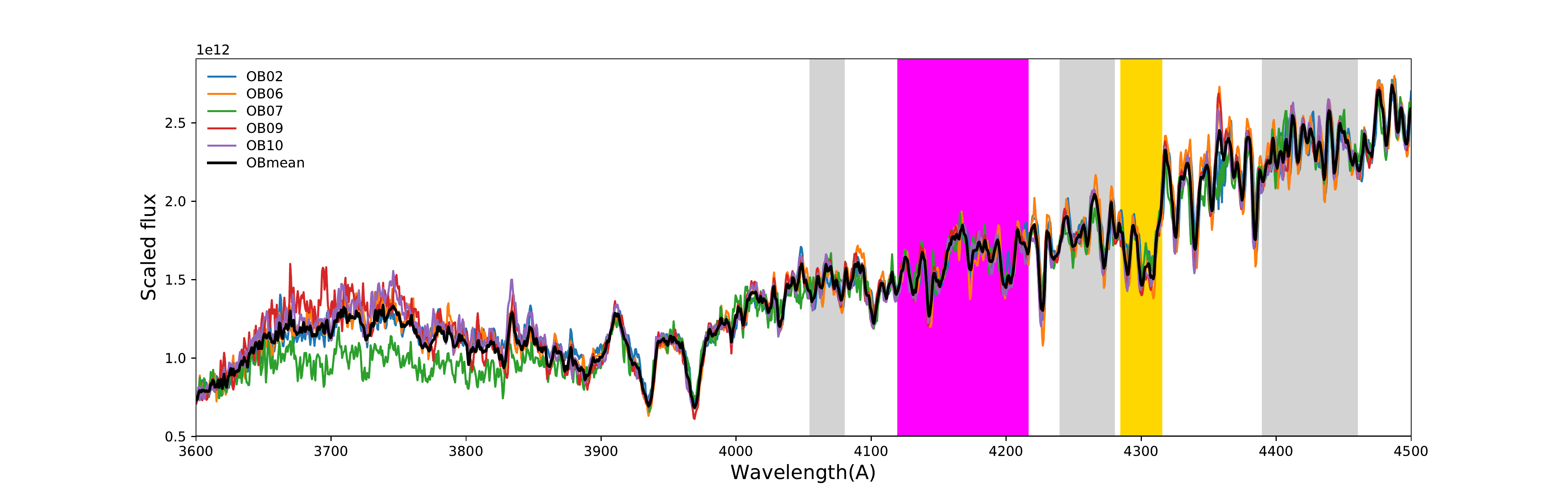}
   \caption{Example of calibrated spectra of five OBs and the co-added one.
            The grey regions indicate the pseudo-continuum, while
            the red and yellow regions indicate the feature wavelengths for measuring the spectral indices of CN4142 and CH4300, respectively.
            }
   \label{Eri-Spec}
  \end{figure}

The data reduction was conducted following the procedure described in \citet{Tang2021}, which can be summarized briefly as follows. The GTCMOS pipeline\footnote{https://www.inaoep.mx/~ydm/gtcmos/gtcmos.html} was first used to complete basic steps: for each OB, the observed spectral images were bias-subtracted, combined, wavelength calibrated, flat-field corrected, and extracted to multiple one-dimension (1D) spectra. After velocity correction, the observed spectra were calibrated using model spectra in order to minimize the uncertainty caused by the flat-field correction.
Considering the stellar parameters estimated in Section \ref{sec:anal-res}, we generated the model spectra with iSpec\footnote{https://www.blancocuaresma.com/s/iSpec} \citep{Blanco2014,Blanco2019}, where the SPECTRUM radiative transfer code and line list, MARCS model atmospheres \citep{Gustafsson2008} and the solar abundances from \citet{Asplund2005} were adopted.
After the calibration, the spectra of the five OBs were co-added for each star (Fig. \ref{Eri-Spec}).

\section{Analysis and results} \label{sec:anal-res}

\subsection{Basic information of stars} \label{subsec:basicInfo}
%---Photometry----
We updated the Gaia photometry with Gaia DR2 data, and extinctions in G, BP and RP bands were computed using the coefficients provided by the team \citep{GaiaDR2}.
We also cross-matched the sample stars with the catalogue of \citet{Munoz2018} to obtain their g and r band magnitudes. For the two stars (eri08 and eri10) that did not get matched, we supplemented their g and r band magnitudes by transforming from the Gaia photometry with the photometric relations provided in the ``Gaia Data Release 2 Documentation release 1.2''\footnote{\scriptsize{https://gea.esac.esa.int/archive/documentation/GDR2/Data\_processing/chap\_cu5pho/sec\_cu5pho\_calibr/ssec\_cu5pho\_PhotTransf.html}}. The systematic differences between the magnitudes of \citet{Munoz2018} and those transformed from Gaia photometry were corrected based on other sample stars. Moreover, V and K band magnitudes were also computed for the sample stars from Gaia photometry. All the photometric magnitudes under consideration are listed in Table \ref{tab:basic}.

\begin{sidewaystable}[thp] \scriptsize
 \caption{Basic information of sample stars.} \label{tab:basic}
 \centering
 \begin{tabular}{c c c c c c c c c c c c c c c c c c}
 \hline
 \hline
  ID  & RA (J2000) & Dec (J2000) & G & e\_G & BP & e\_BP & RP & e\_RP &  g & e\_g & r & e\_r &
  pmRA & e\_pmRA & pmDec & e\_pmDec & Evol. phase    \\
      &          &           & (mag) & (mag) & (mag) & (mag) & (mag) & (mag) & (mag) & (mag) & (mag) &
  (mag) & (mas/yr) & (mas/yr) & (mas/yr) & (mas/yr) &     \\
 \hline
eri01  &    66.176910  &  -21.167276  &    19.650  &   0.005  &   19.992  &   0.054  &   18.885  &   0.030  &    20.215  &   0.003  &   19.560  &   0.003  &    1.1810  &   0.6248  &   -0.8417  &   0.6942  &   RGB   \\
eri02  &    66.168472  &  -21.170403  &    20.322  &   0.007  &   20.602  &   0.083  &   19.694  &   0.047  &    20.809  &   0.004  &   20.194  &   0.003  &   -0.0457  &   1.0429  &    1.4722  &   1.2958  &   RGB   \\
eri03  &    66.182945  &  -21.174185  &    18.013  &   0.002  &   18.687  &   0.018  &   17.251  &   0.009  &    18.872  &   0.002  &   17.946  &   0.002  &    0.7988  &   0.2064  &   -0.1627  &   0.2227  &   RGB   \\
eri04  &    66.185150  &  -21.175451  &    18.340  &   0.002  &   18.994  &   0.025  &   17.593  &   0.011  &    19.125  &   0.002  &   18.267  &   0.002  &    0.3653  &   0.2587  &   -0.9767  &   0.2822  &   RGB   \\
eri05  &    66.173363  &  -21.179552  &    18.407  &   0.002  &   18.984  &   0.021  &   17.652  &   0.014  &    19.178  &   0.002  &   18.332  &   0.002  &    0.5388  &   0.2920  &   -0.4840  &   0.3129  &   RGB   \\
eri06  &    66.194931  &  -21.180887  &    19.862  &   0.005  &   20.070  &   0.048  &   19.231  &   0.062  &    20.263  &   0.003  &   19.782  &   0.003  &   -0.5298  &   0.8228  &    0.4895  &   0.7989  &    HB   \\
eri07  &    66.184982  &  -21.181849  &    18.773  &   0.003  &   19.252  &   0.028  &   18.107  &   0.020  &    19.390  &   0.002  &   18.696  &   0.002  &    0.7575  &   0.3397  &   -0.5467  &   0.3856  &   AGB   \\
eri08  &    66.183861  &  -21.183519  &    17.328  &   0.002  &   17.876  &   0.012  &   16.388  &   0.006  &    18.309  &   0.000  &   17.248  &   0.000  &    0.4355  &   0.1299  &   -0.1746  &   0.1437  &   RGB   \\
eri09  &    66.181190  &  -21.184744  &    18.068  &   0.002  &   18.680  &   0.020  &   17.290  &   0.009  &    18.895  &   0.002  &   17.999  &   0.002  &    0.7367  &   0.2031  &   -1.0208  &   0.2285  &   RGB   \\
eri10  &    66.187309  &  -21.185621  &    17.073  &   0.002  &   17.835  &   0.010  &   16.203  &   0.005  &    18.210  &   0.000  &   17.020  &   0.000  &    0.4044  &   0.1173  &   -0.5422  &   0.1333  &   RGB   \\
eri11  &    66.188339  &  -21.188213  &    19.143  &   0.003  &   19.450  &   0.035  &   18.369  &   0.025  &    19.691  &   0.003  &   19.050  &   0.003  &    0.5976  &   0.4182  &   -0.9497  &   0.4862  &   AGB   \\
eri12  &    66.195686  &  -21.189545  &    20.644  &   0.008  &   20.757  &   0.122  &   19.920  &   0.090  &    21.113  &   0.004  &   20.531  &   0.004  &    3.4233  &   1.5259  &   -0.2365  &   1.3503  &   RGB   \\
eri13  &    66.185699  &  -21.190716  &    20.558  &   0.009  &   20.512  &   0.082  &   19.853  &   0.095  &    20.997  &   0.005  &   20.415  &   0.004  &   -0.2745  &   1.3342  &    2.3632  &   1.5846  &   RGB   \\
eri14  &    66.182388  &  -21.191929  &    20.171  &   0.007  &   20.453  &   0.070  &   19.597  &   0.057  &    20.500  &   0.004  &   20.080  &   0.003  &    0.5595  &   0.9482  &   -1.7182  &   1.2901  &    HB   \\
eri15  &    66.185364  &  -21.195387  &    18.719  &   0.003  &   19.251  &   0.020  &   18.016  &   0.015  &    19.426  &   0.002  &   18.637  &   0.002  &    0.8461  &   0.3807  &   -0.6992  &   0.3909  &   RGB   \\
eri16  &    66.186218  &  -21.197334  &    20.149  &   0.006  &   20.312  &   0.050  &   19.496  &   0.038  &    20.493  &   0.003  &   20.081  &   0.003  &    0.0457  &   0.9189  &    2.1428  &   1.0991  &    HB   \\
eri17  &    66.177940  &  -21.202587  &    18.988  &   0.003  &   19.528  &   0.025  &   18.262  &   0.019  &    19.642  &   0.002  &   18.903  &   0.002  &    0.3619  &   0.3844  &    0.0074  &   0.4424  &   RGB   \\
eri18  &    66.172325  &  -21.141823  &    20.047  &   0.007  &   20.092  &   0.065  &   19.488  &   0.094  &    20.232  &   0.003  &   19.972  &   0.004  &   -1.3608  &   0.6670  &   -3.3207  &   0.8347  &    HB   \\
eri19  &    66.189949  &  -21.221462  &    20.678  &   0.010  &   20.901  &   0.103  &   20.239  &   0.115  &    21.012  &   0.004  &   20.581  &   0.005  &    2.8056  &   1.6933  &   -2.7386  &   2.0293  &    HB   \\
 \hline
 \end{tabular}
\end{sidewaystable}

\begin{figure}[ht]
   \centering
     \includegraphics[width=0.46\textwidth]{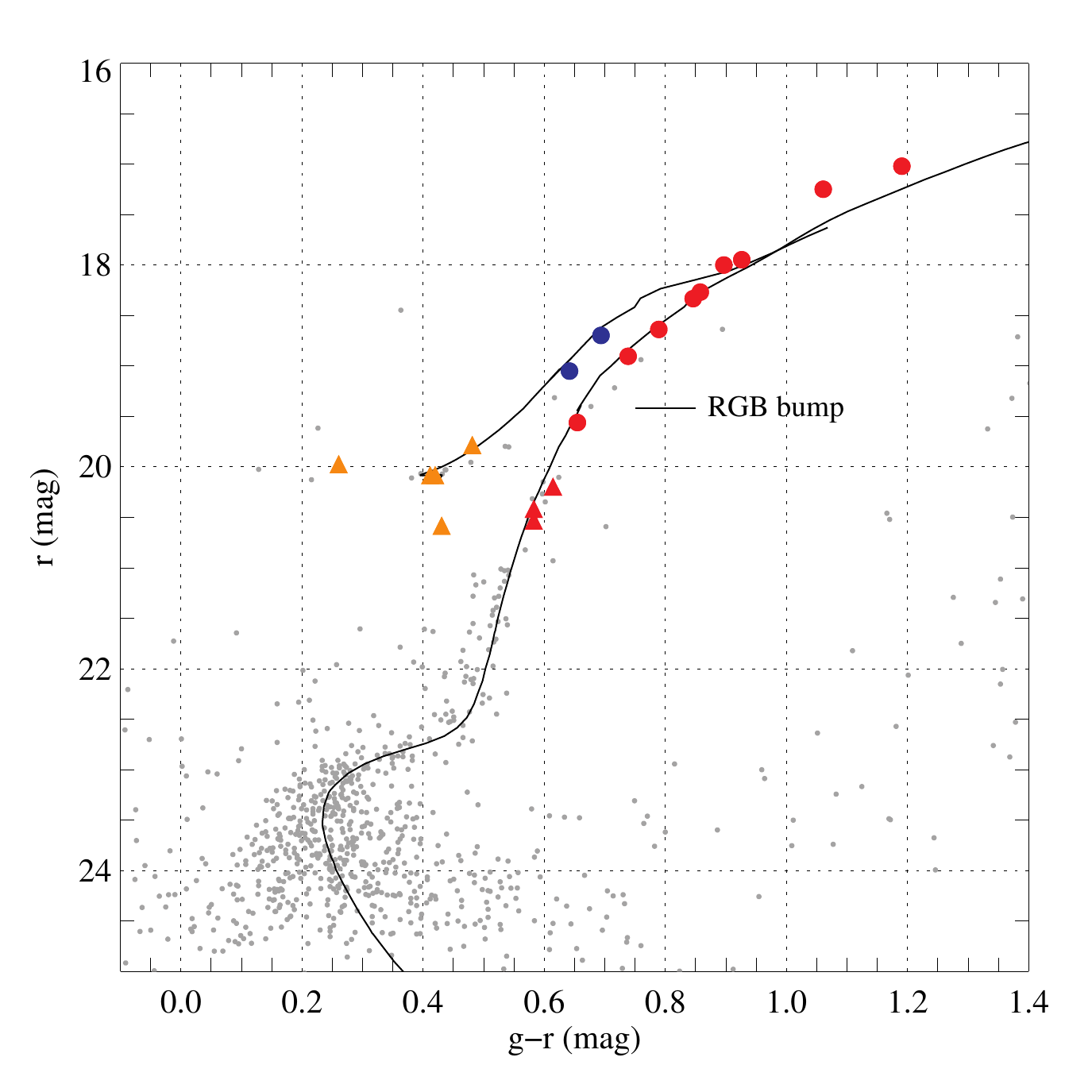}
   \caption{The color$-$magnitude diagram of Eridanus. The colors red, blue and yellow represent RGB, AGB and HB stars in our sample, respectively.
            The solid circles and triangles mark the status of the stellar proper motions (see the text for details).
            Grey dots are stars from \citet{Munoz2018} that are located within the tidal radius of $3.17'$ \citep{Myeong2017}.
            A PARSEC isochrone of the age of 10\,Gyr and metallicity of -1.45 is shown, and the position of the RGB bump is also indicated.
            }
   \label{fig:Eri-CMD}
  \end{figure}

%\newpage{}
%---CMD and pm---
Figure \ref{fig:Eri-CMD} shows the color$-$magnitude diagram (CMD) of Eridanus, where the photometric catalogue of \citet{Munoz2018} is used. Grey dots indicate stars located within the tidal radius of $3.17'$ \citep{Myeong2017}.
Using grids of PARSEC isochrones to fit the CMD distribution, we determined the age of Eridanus to be 10.0\,$\pm$\,0.5\,Gyr and metallicity to be -1.45\,$\pm$\,0.05\,dex.
The distance modulus and reddening E(g-r) were also derived as 19.83\,$\pm$\,0.04\,mag and 0.025\,$\pm$\,0.010\,mag, respectively.
A distance of 92.47\,$\pm$\,1.71\,kpc is then obtained based on the distance modulus.
Considering the V-band magnitude 15.02\,mag of Eridanus derived by \citet{Baumgardt2020}, the absolute visual magnitude is $M_{V}$\,=\,-4.81\,mag.
E(B-V) of 0.021\,$\pm$\,0.008\,mag was derived adopting the extinction coefficients by \citet{McCall2004}.
These parameters have also been determined in previous studies.
For example, the metallicity of -1.35$\pm$0.2\,dex, distance of 81\,kpc to the Sun and absolute visual magnitude of -4.85\,mag have been derived by \citet{DaCosta1985} via CMD fitting for the first time.
The distance of 90.1\,kpc to the Sun by \citet{Stetson1999} and metallicity of -1.43 which is the average of the values by \citet{Armandroff1991} and \citet{Carretta2009} are adopted by \citet[][2010 edition]{Harris1996} in the GC catalogue.
\citet{Baumgardt2021} derived a distance of 84.68\,kpc.
An absolute age of 10.5\,Gyr was obtained by \citet{Beccari2012} for this GC.
For E(B-V), a weighted average value of 0.02\,mag considering several sources is adopted by \citet[][2010 edition]{Harris1996}, and \citet{Munoz2018} determined it to be 0.018\,mag.
We conclude that our results of these parameters are consistent with those of previous works.

\begin{figure}[ht]
   \centering
     \includegraphics[width=0.46\textwidth]{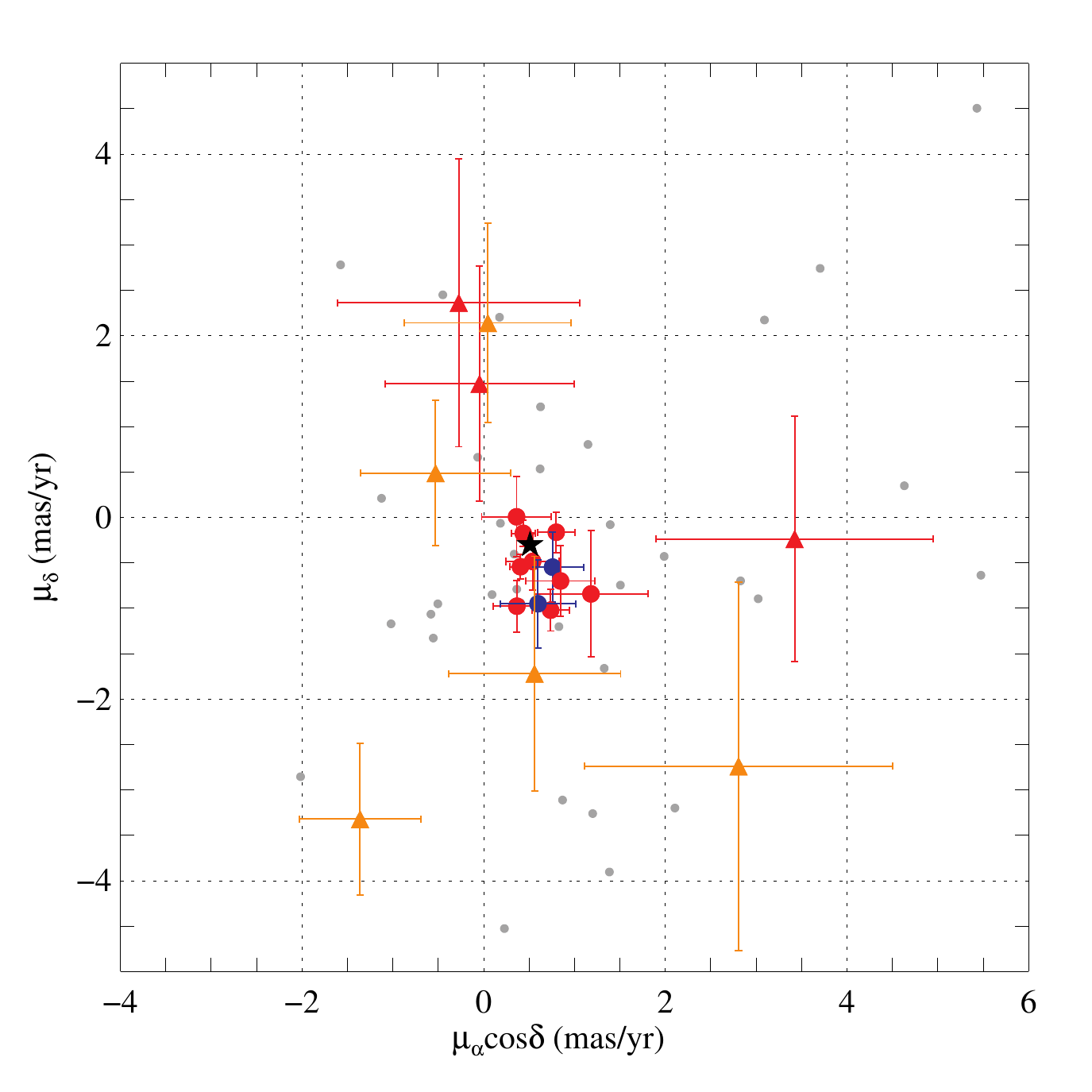}
   \caption{Proper motions of our sample stars. The symbols are the same as in Fig. \ref{fig:Eri-CMD} and the size of error bars indicates the uncertainties in proper motions. Solid triangles represent the HB stars and the faintest three RGB stars, which have much larger errors in their proper motions compared with the brighter RGB and AGB stars, represented by solid circles. Grey dots are part of the background stars shown in Fig. \ref{fig:Eri-CMD} that have Gaia DR2 proper motion values. The black star indicates the mean proper motion of Eridanus derived by \citet{Vasiliev2021}.
            }
   \label{fig:Eri-pm}
  \end{figure}

The stars that we observed are shown in Fig. \ref{fig:Eri-CMD} with large colored symbols. There are two AGB stars, 12 RGB stars\footnote{\scriptsize{For two of these stars, it is difficult to distinguish between RGB or AGB based on their positions in CMD, but in our study we take them as RGB stars.}} and five HB stars (Table \ref{tab:basic}).
We checked their proper motions from Gaia DR2 \citep{GaiaDR2} as shown in Fig. \ref{fig:Eri-pm}.
All the HB stars and the faintest three RGB stars (represented by solid triangles in Figs. \ref{fig:Eri-CMD} and \ref{fig:Eri-pm}) have much larger uncertainties in their proper motions, compared with those brighter RGB and AGB stars (represented by solid circles). Since the G magnitudes of the fainter ones in our sample are about 20-21 mag, their uncertainties are comparable to the documented typical values (1.2-3 mas yr$^{-1}$) at this magnitude range \citep{Brown2018}.

\subsection{Stellar parameters}
%---Stellar parameter---
The effective temperature ($T_\mathrm{eff}$) was first determined based on the Gaia photometry, considering the color$-$temperature relations provided in \citet{Mucciarelli2020}. Six relatively independent determinations of $T_\mathrm{eff}$ were computed based on the de-reddened colors of (BP-RP), (BP-G), (G-RP), (BP-K), (RP-K) and (G-K).
Converting the de-reddened (g-r) color into the color (B-V) adopting the transformation provided by \citet{Jester2005}, we also considered another group of $T_\mathrm{eff}$ based on the color (B-V), which was derived using the color$-$temperature relation provided in \citet{GH2009}.
The final $T_\mathrm{eff}$ was determined as the mean value of all the above seven values of $T_\mathrm{eff}$ after a sigma clipping.
The associated errors were estimated taking the dispersions of the photometric calibrations and the uncertainties in the colour index and [Fe/H] into account.
We estimated the stellar masses based on the isochrone (Table \ref{tab:SPAbu}). Then, the surface gravities ($\log g$) were derived from $T_\mathrm{eff}$, stellar masses, and bolometric luminosities. The relations of \citet{Alonso1999} were used to calculate the bolometric corrections.

Considering the derived $T_\mathrm{eff}$ of HB stars and the three faintest RGB stars (represented by solid triangles in Figs. \ref{fig:Eri-CMD} and \ref{fig:Eri-pm}) are all above 5500\,K, the CN and CH molecular lines are weak for these hot stars; moreover, their uncertainties in proper motions are much larger than other brighter stars and their SNRs are substantially lower, we opted to neglect these hotter and fainter stars in the following analysis and discussion. In other words, we only consider the brighter RGB and AGB stars as our final sample.
The derived $T_\mathrm{eff}$, $\log g$ and their associated errors for each star in the final sample are listed in Table \ref{tab:SPAbu}.
The typical errors on $T_\mathrm{eff}$ and $\log g$ are about $\pm$50\,K and $\pm$0.03\,dex, respectively.

\subsection{Spectral indices and abundances}
%---CH CN indicies---
Following the definition of \citet{Harbeck2003}, we computed the spectral indices of CN4142 and CH4300 on the co-added calibrated spectra for each star.
\begin{equation}
{\rm CN4142}=-2.5\log\frac{F_{4120-4216}}{0.4F_{4055-4080}+0.6F_{4240-4280}}
\end{equation}
\begin{equation}
{\rm CH4300}=-2.5\log\frac{F_{4285-4315}}{0.5F_{4240-4280}+0.5F_{4390-4460}}
\end{equation}
where $F_{X-Y}$ is the summed spectral flux from X to Y \AA.
Figure \ref{Eri-Spec} shows the spectral regions for measuring the indices of CN4142 and CH4300.
The errors of spectral indices were estimated mainly considering two factors, i.e., the errors of calibrating observed spectra with model spectra and the flux dispersion among calibrated OB spectra.
Table \ref{tab:SPAbu} lists the derived spectral index values and their errors.
Note that we did not plot $\delta$CN versus $\delta$CH \citep[e.g.,][]{Gerber2020} here, because we do not have enough stars to define base lines where $\delta$CN$=0$ and $\delta$CH$=0$.

  \begin{figure}[ht]
   \centering
     \includegraphics[width=0.6\textwidth]{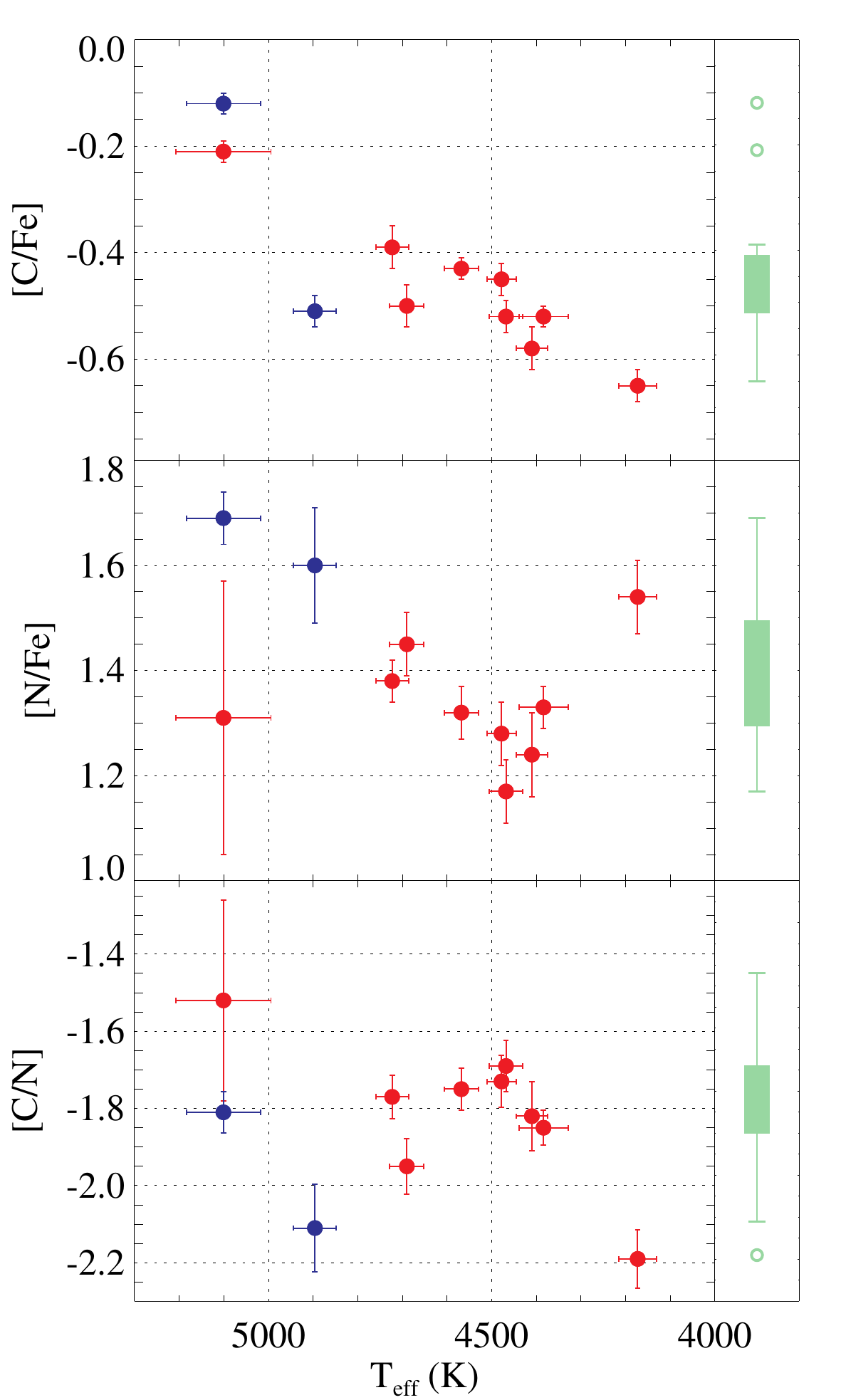}
   \caption{Distribution of the abundance ratios of [C/Fe], [N/Fe] and [C/N].
            Symbol colors are the same as in Fig. \ref{fig:Eri-CMD}. The box-plot of each abundance distribution is also shown on the right side of each panel, where the green thick bar indicates the interquartile range (IQR) covering the central 50\% of the data, the thin lines extending from the thick bar covering the range of the remaining data, while the individual small circles mark the outliers located further than 1.5 times the IQR from the edge of the thick bar.
            }
   \label{fig:AbuCNFe}
  \end{figure}

To analyze the [C/Fe] and [N/Fe] specifically, we took advantage of the model spectra from iSpec. For each star, we generated a grid of model spectra with [C/Fe$] = -0.8, -0.7, -0.6, ... 0$ and [N/Fe$] = 0.3, 0.4, 0.5, ... 1.8$.  [O/Fe]  was assumed to be solar for all the model spectra.\footnote{ According to our test shown in \citet{Tang2021}, the index differences for CN4142 and CH4300 caused by an [O/Fe] difference of 0.4\,dex  is almost negligible inside the discussed temperature range. }
The model spectral indices of CN4142 and CH4300 were computed for each model spectrum, so that the grid of model spectral indices was constructed for each star.
By interpolation, we found the best match to the observed spectral indices from the model grid, so that a pair of best-match [C/Fe] and [N/Fe] was derived. In order to derive more reliable abundances as well as to estimate their errors, we performed the interpolation 1000 times for each star, considering $1\sigma$ variations in a Gaussian Monte Carlo approach in the input data. The mean abundances and their standard deviations are adopted as the final abundances of [C/Fe] and [N/Fe] and their errors, where the typical errors (median value) are 0.03\,dex and 0.06\,dex for [C/Fe] and [N/Fe], respectively. The results of the final sample stars are summarized in Table \ref{tab:SPAbu}.
By examining the dispersions ($\sigma$) and the interquartile range (IQR) values of carbon and nitrogen abundances, which are $\sigma_{[C/Fe]}$\,=\,0.15, IQR$_{[C/Fe]}$\,=\,0.11 and $\sigma_{[N/Fe]}$\,=\,0.16, IQR$_{[N/Fe]}$\,=\,0.20, respectively, we find that the dispersions of [C/Fe] and [N/Fe] are similar and both significantly larger than their associated errors, while [C/Fe] distributes more concentratedly.
These can also be detected visually in Fig. \ref{fig:AbuCNFe}, which shows the distributions of [C/Fe] and [N/Fe], as well as the [C/N] ratio.

\begin{table}[thp] \scriptsize
 \caption{Stellar parameters, spectral indices and abundances of the final sample stars.} \label{tab:SPAbu}
 \centering
 \begin{tabular}{c c c c c c c c c c c c c c}
 \hline
 \hline
 ID  &  mass  & $T_\mathrm{eff}$ & e\_$T_\mathrm{eff}$ & $\log g$ & e\_$\log g$ &
 CN4142 & e\_CN4142 & CH4300 & e\_CH4300 & [C/Fe] & e\_[C/Fe] & [N/Fe] & e\_[N/Fe]    \\
    & ($M_{\odot}$) &  (K)  &  (K)  &   &   &   &   &   &   &   &   &   &     \\
 \hline
eri01  &  0.847  &  5101  &  106.58  &   2.15  &  0.05  &  -0.017  &   0.007  &   0.209  &   0.005  &  -0.21  &   0.02  &   1.31  &   0.26   \\
eri03  &  0.836  &  4410  &   35.14  &   1.18  &  0.03  &  -0.005  &   0.008  &   0.273  &   0.007  &  -0.58  &   0.04  &   1.24  &   0.08   \\
eri04  &  0.839  &  4468  &   37.62  &   1.34  &  0.03  &  -0.012  &   0.006  &   0.279  &   0.006  &  -0.52  &   0.03  &   1.17  &   0.06   \\
eri05  &  0.840  &  4568  &   38.66  &   1.42  &  0.03  &   0.007  &   0.008  &   0.281  &   0.005  &  -0.43  &   0.02  &   1.32  &   0.05   \\
eri07  &  0.780  &  4896  &   47.80  &   1.67  &  0.03  &   0.005  &   0.007  &   0.185  &   0.005  &  -0.51  &   0.03  &   1.60  &   0.11   \\
eri08  &  0.828  &  4384  &   55.01  &   0.89  &  0.04  &   0.020  &   0.006  &   0.290  &   0.005  &  -0.52  &   0.02  &   1.33  &   0.04   \\
eri09  &  0.837  &  4478  &   32.70  &   1.24  &  0.03  &   0.007  &   0.007  &   0.290  &   0.006  &  -0.45  &   0.03  &   1.28  &   0.06   \\
eri10  &  0.819  &  4173  &   42.12  &   0.67  &  0.03  &   0.037  &   0.008  &   0.274  &   0.006  &  -0.65  &   0.03  &   1.54  &   0.07   \\
eri11  &  0.790  &  5101  &   83.02  &   1.91  &  0.04  &   0.006  &   0.005  &   0.214  &   0.005  &  -0.12  &   0.02  &   1.69  &   0.05   \\
eri15  &  0.842  &  4723  &   36.71  &   1.61  &  0.03  &  -0.001  &   0.004  &   0.264  &   0.010  &  -0.39  &   0.04  &   1.38  &   0.04   \\
eri17  &  0.844  &  4690  &   38.31  &   1.71  &  0.03  &   0.006  &   0.007  &   0.245  &   0.010  &  -0.50  &   0.04  &   1.45  &   0.06   \\
 \hline
 \end{tabular}
\end{table}

  \begin{figure}[ht]
   \centering
     \includegraphics[width=0.55\textwidth]{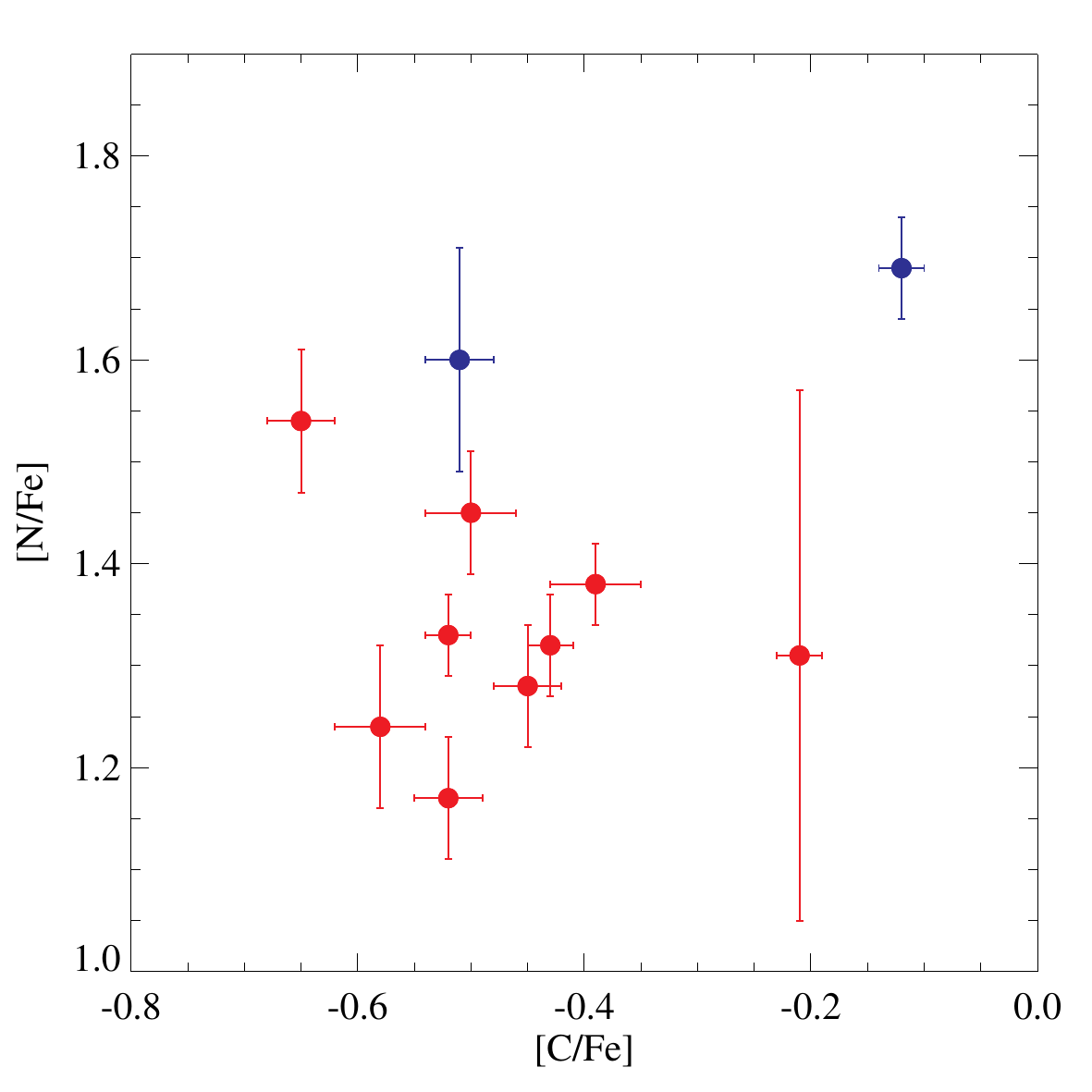}
   \caption{ [C/Fe] {\it vs.} [N/Fe]. Symbols are the same as in Fig. \ref{fig:Eri-CMD}.
            }
   \label{fig:CFeNFe}
  \end{figure}

The activated CNO cycle in the cores of evolved low-mass stars may change the surface chemical abundances if the convective envelope reaches deep enough. As stars climb up the RGB, the surface abundances of carbon and nitrogen would be altered due to the first dredge-up and extra-mixing processes. Especially, the extra-mixing process can significantly change the surface abundances of stars brighter than the RGB bump \citep[e.g.,][]{Gratton2000,Charbonnel2007,Lagarde2012}.
As shown in Fig. \ref{fig:Eri-CMD}, only the faintest RGB star in our final sample is located below the RGB bump, while all the others should have experienced the extra-mixing. So the abundance distributions shown in Fig. \ref{fig:AbuCNFe} are mixture results of the original chemical composition and their self-evolution.  In that sense, the lack of anti-correlation for [C/Fe] $\it vs.$ [N/Fe] (Fig. \ref{fig:CFeNFe}) is not surprising. Similar situation can be found for other GCs with small sample size \citep[e.g.,][]{Meszaros2020}.

\begin{figure}[ht]
   \centering
     \includegraphics[width=0.6\textwidth]{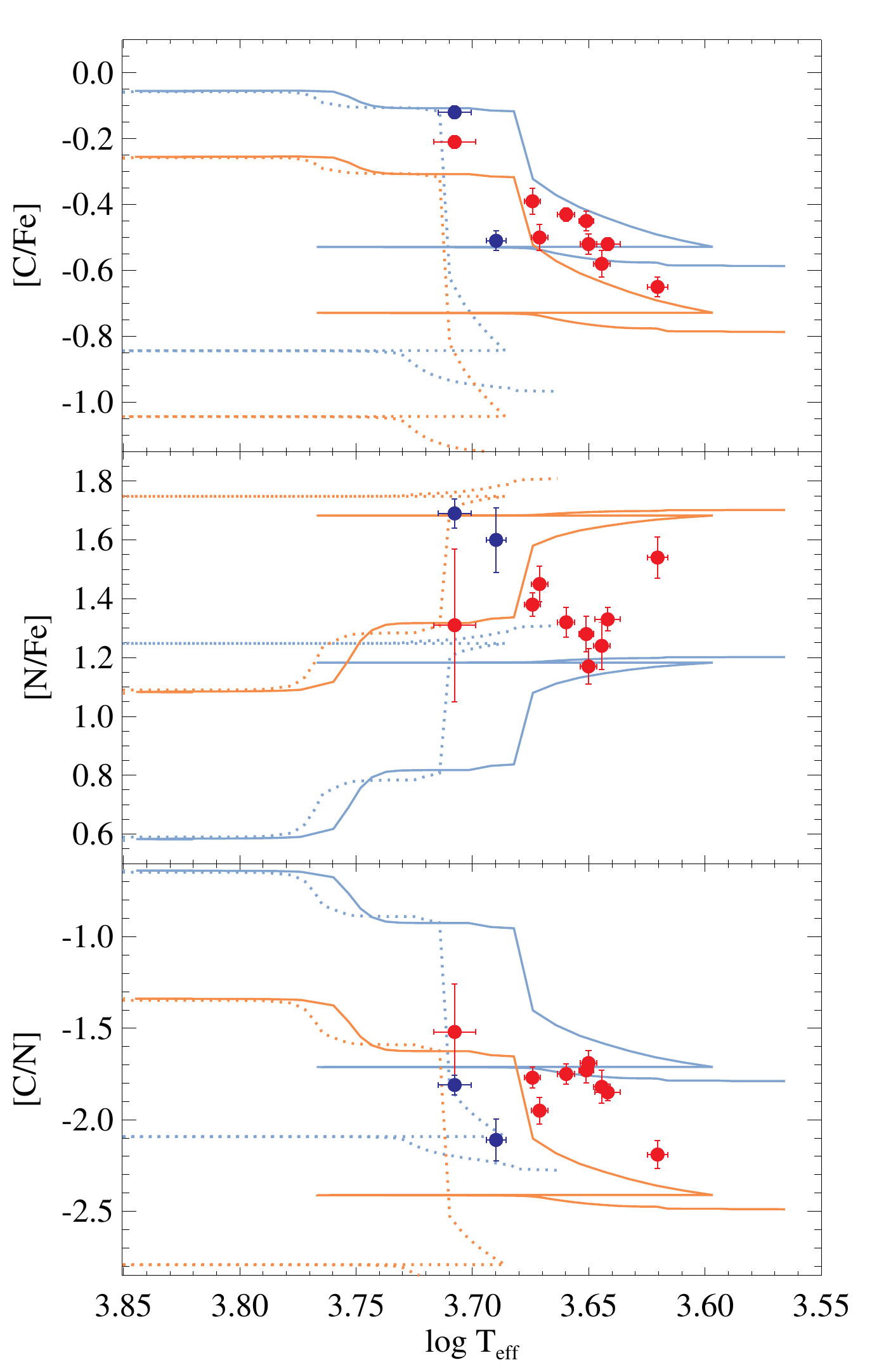}
   \caption{Comparison between the distributions of [C/Fe], [N/Fe], [C/N] and the model predictions.  Symbols are the same as in Fig. \ref{fig:Eri-CMD}.
           Models including thermohaline convection and rotation-induced mixing from \citet{Lagarde2012} are considered here. Blue and orange lines represent model predictions for the stellar mass of 0.85\,$M_{\odot}$ with initial chemical compositions of [C/Fe]\,=\,-0.05, [N/Fe]\,=\,+0.6 and [C/Fe]\,=\,-0.25, [N/Fe]\,=\,+1.1, respectively, mimicking the primordial and enriched stellar populations, while solid and dotted lines represent those for metallicities of Z\,=\,0.002 and 0.0001 ([Fe/H]\,=\,-0.86 and -2.16), respectively.
            }
   \label{fig:AbuCNFeModel2}
  \end{figure}

To investigate the influence of the extra-mixing process in our sample, we took advantage of models of \citet{Lagarde2012} where thermohaline convection and rotation-induced mixing are included. In parallel with our derived abundances, the model predictions for the stellar mass of 0.85\,$M_{\odot}$ are shown in Fig. \ref{fig:AbuCNFeModel2}, considering the initial chemical compositions of [C/Fe]\,=\,$-0.05$, [N/Fe]\,=\,$+0.6$ (blue lines) and [C/Fe]\,=\,$-0.25$, [N/Fe]\,=\,$+1.1$ (orange lines), mimicking the primordial and enriched stellar populations, respectively.
In term of carbon abundances, the observed data points are well constrained by models with [C/Fe]\,=\,$-0.25 $ --- $ -0.05$, and the abundance trend with $T_\mathrm{eff}$ follows exactly the model prediction. This implies that: a) the decrease in carbon abundance with decreasing $T_\mathrm{eff}$ is obviously a result of stellar evolutionary mixing; b) the initial carbon abundance spread among our sample stars is around $\sim$0.2\,dex, which is slightly larger than our determination error ($\sim$\,0.03\,dex in average).
On the other hand, the surface abundance of nitrogen is expected to increase as a result of extra-mixing, but the observed values do not follow the predictions of models with a single or a narrow range of initial [N/Fe]. Clearly, multiple models with initial [N/Fe] between +0.6 and +1.1 are needed to cover our data points. A difference of 0.5\,dex in [N/Fe] is no doubt larger than the observed abundance errors ($\sim$\,0.08\,dex in average).
Moreover, a scatter of about 0.7\,dex in [C/N] can also be deduced, which is also significant compared with the error of $\sim$\,0.06\,dex in average.
Therefore, the existence of abundance variations among our observed member stars are suggested by (1) the apparent abundance distributions with no evolutionary effect considered (Fig. \ref{fig:AbuCNFe}) and (2) the stellar evolutionary models considering extra-mixing processes of \citet{Lagarde2012} (Fig. \ref{fig:AbuCNFeModel2}).
In conclusion, we find clear evidence that indicates the existence of MPs in GC Eridanus, particularly in nitrogen abundances.

\section{Discussion} \label{sec:disc}

\subsection{Eridanus}

 Given its higher metallicity ([Fe/H$]\sim -1.4$) compared to other outer halo GCs, and slightly younger age (10.5 Gyr), Eridanus is considered to have formed ex-situ.
\citet{CarballoBello2015} speculated that Eridanus may be related to the Monoceros Ring since this GC lies in its predicted orbit. However, recent studies showed that the Monoceros Ring to be part of the outer Galactic disk with Galactocentric distances $< 30$ kpc \citep{LiJ2021}, and therefore their connection seems unlikely. On the other hand, \citet{Myeong2017} discovered two symmetrical tidal tails around Eridanus using g- and i-band deep imaging, which span about 760\,pc in length. According to the great circle generated along the direction of the tidal tails, they instead suggested that Eridanus could have possible association with one of the four dwarf galaxies: Sculptor, Canes Venatici\,I, Canes Venatici\,II, and Fornax, of which Sculptor has a similar distance as Eridanus.
The discovery of symmetrical, long tidal tails around Eridanus indicates that it has gone through a significant mass loss phase. This has two implications: 1. the initial mass should be much higher than the current mass, and thus our discovery of MPs in this GC does not necessarily contradict the current MP  scenarios. A similar situation was found in Pal 13 \citep{Tang2021}; 2. Eridanus could have undergone stripping due to a radial orbit, causing stars to drift out from the Lagrange points of the cluster. \,\, However, its connection with low mass dwarf galaxies, e.g., Sculptor, may cause a logical dilemma. The existence of Eridanus is at odds with the lack of existing GC in Sculptor, and our discovery of N-enhanced stars in Eridanus is inconsistent with the null-detection of N-rich field stars in Sculptor \citep{Lardo2016, Tang2022}. Therefore, their association is less likely from a chemical point of view.
\citet{Massari2019} analyzed the origins of Galactic clusters combining kinematic information and available cluster ages, where Eridanus is suggested to be accreted from a low-mass progenitor. There are still many unanswered questions regarding its origin, evolution, composition, and etc. The upcoming China Space Station Telescope (CSST) will give us a more complete picture of Eridanus by: 1. exploring its tidal tail mass with deep photometry; 2. investigating its cluster mass function and multiple populations on the main sequence with UV-optical filters.

\subsection{Threshold mass for MP formation}

\begin{figure}[ht]
   \centering
     \includegraphics[width=0.7\textwidth]{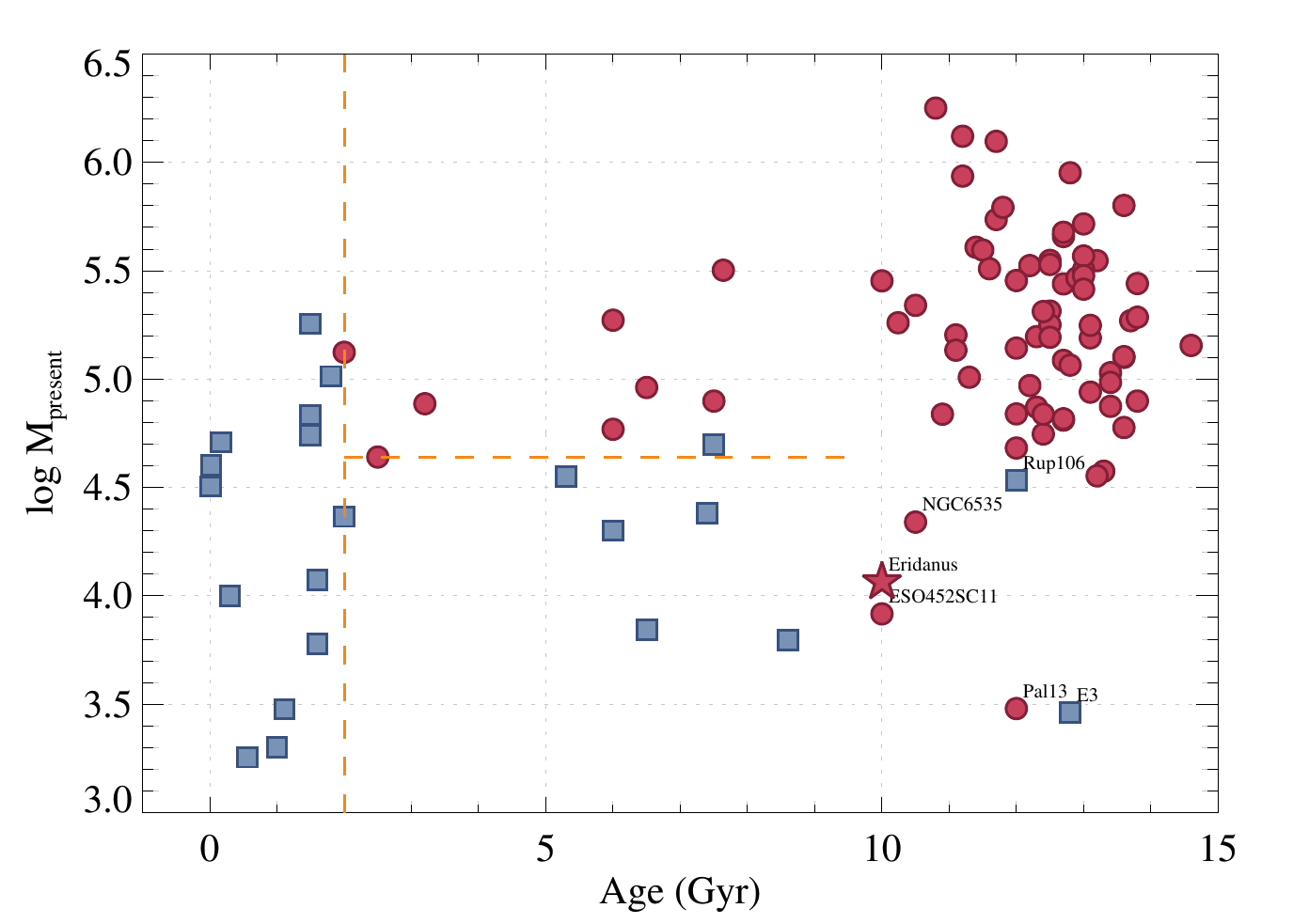}
   \caption{Age-present mass plane for clusters with (red circles) and without (blue squares) MPs. Eridanus is presented with a red star. The orange dashed lines indicate the rough separation in age and mass between clusters with and without MPs.}
   \label{fig:AgeCMass}
  \end{figure}

The distribution of age $vs.$ present-day mass for clusters with and without MPs was initially presented by \citet{Martocchia2018}, where the data of the compilation by \citet{Krause2016} was considered.
We further add data points from recent works, e.g.,
\citet{Salinas2015,Niederhofer2017a,Niederhofer2017b,Simpson2017,Tang2017,Hollyhead2018,Zhang2018,Li2019a,Li2019b,Li2020,Milone2020,Tang2021,FT2022}, as well as this work (Fig. \ref{fig:AgeCMass}).
We take the masses of the Milky Way GCs from the most recent version of the Galactic Globular Cluster Database from \citet{Baumgardt2018} which has been updated to include the Gaia EDR3 data. We also include data for several large clusters from the Large and Small Magellanic Clouds (LMC/SMC). The ages of the clusters have been taken from recent literature while their present-day masses are calculated by fitting N-body models to archival HST photometry, similar to what \citet{Baumgardt2018} have done for Galactic GCs. The initial masses for these clusters are calculated by using the present-day masses and applying the models of \citet{Baumgardt2013}.

As previously suggested, the separation of clusters with and without MPs at the age of $\sim$2\,Gyr can be clearly seen in the figure. For clusters younger than 10\,Gyr, those with masses under $\log M \sim$ 4.65 do not show MPs.
However, the condition is ambiguous for GCs over 10\,Gyr, since several low-mass GCs, e.g. NGC\,6535 \citep{Bragaglia2017}, Eridanus (this work, represented by a red star), ESO452-SC11 \citep{Simpson2017}, Pal\,13 \citep{Tang2021}, also possess MPs. According to the updated mass estimation, Pal\,13 is the lowest-mass cluster with MP signals. The studies of old, low-mass GCs play a key role in revealing the MP formation and evolution.

\begin{figure}[ht]
   \centering
     \includegraphics[width=0.7\textwidth]{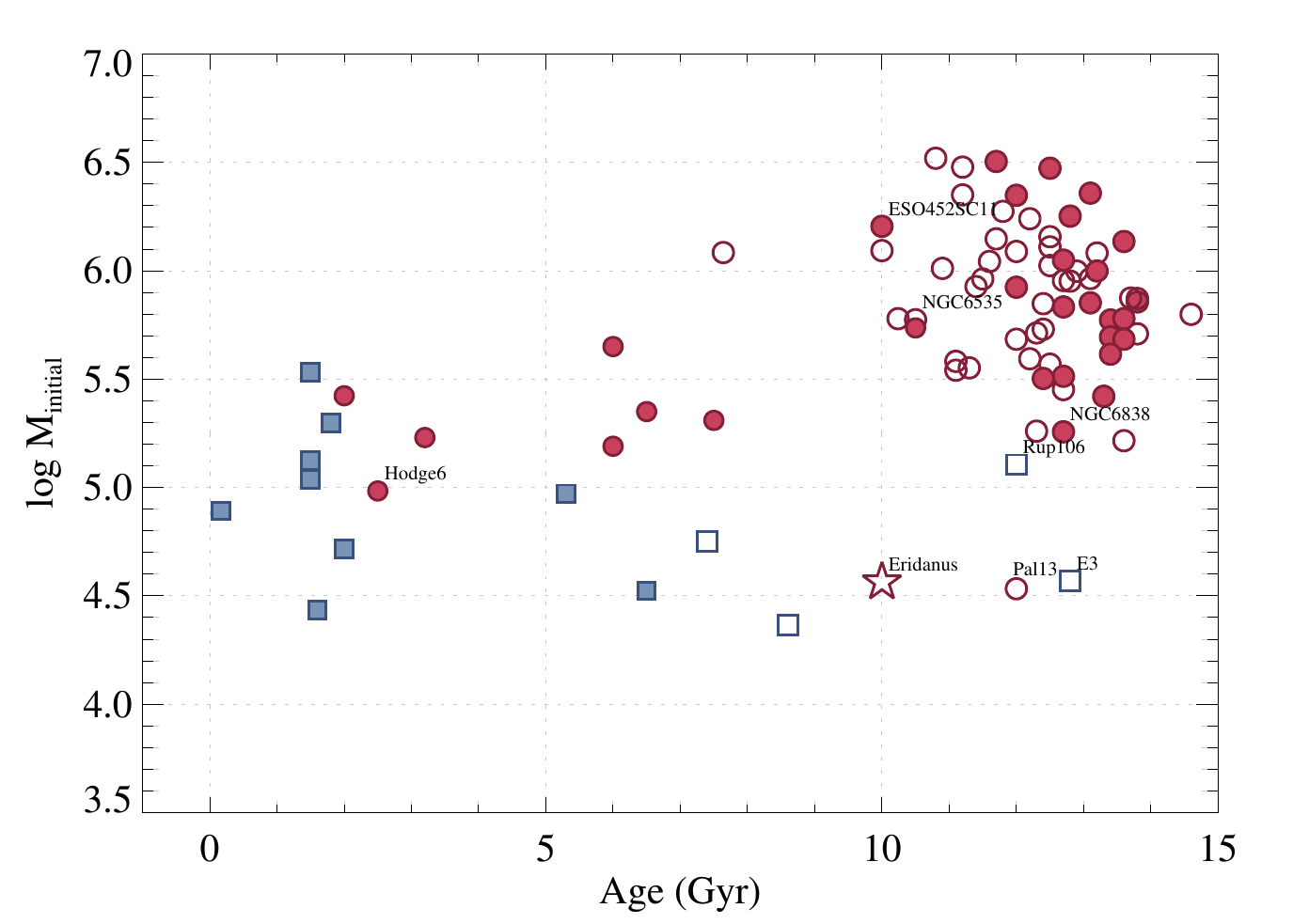}
   \caption{Age-initial mass plane for clusters with and without MPs.  The symbol shapes and colors are the same as in Fig. \ref{fig:AgeCMass}. The origins of the clusters are indicated according to \citet{Massari2019}, where solid symbols represent in-situ clusters, and open ones represent accreted clusters.}
   \label{fig:AgeIMass}
  \end{figure}

During the long-time self evolution and co-evolution with MW,  old GCs may experience complex mass loss. If a GC has an initial mass high enough to form MPs, and then loses a large amount of its mass during its long-time evolution, it can become an old, low-mass GC showing MPs.
In this context, the information of cluster initial masses is fundamental to understand the MP formation \citep[e.g.,][]{Carretta2010,Milone2020}.
\citet{Baumgardt2019} computed the initial masses for Galactic GCs using N-body simulation, while those of clusters in LMC/SMC will be presented in a forthcoming paper.
In Fig. \ref{fig:AgeIMass} we present the age-initial mass distribution of the cluster sample, where the LMC/SMC clusters without known initial masses are excluded.
We find that the dispersion in initial cluster masses is smaller compared to that of their present-day masses, and most old (age $\geq$ 10\,Gyr) GCs with MPs have initial masses above $\log M_{initial}\sim$ 5.2. Two of the aforementioned old low-mass GCs, NGC\,6535 and ESO452-SC11, have initial masses as high as most of other old GCs, and thus their MPs can be easily explained.
For Eridanus and Pal\,13, considering their probable external origins, their real initial masses could be higher, since \citet{Baumgardt2019} only considered the GC orbits within the Milky Way in the simulation, and the mass loss before and during their merger into the Milky Way have been omitted.

Considering uncertainties in the initial mass estimation for accreted clusters, in-situ ones can provide a simpler sample for analyzing the cluster mass threshold for MP formation.
We divide the Galactic GCs into in-situ ones and accreted ones, according to category in  \citet{Massari2019}. Clusters in LMC and SMC are treated as in-situ ones (Fig. \ref{fig:AgeIMass}).
In the in-situ sample, there is a slight trend in the clusters with MPs (older than $\sim$ 2\,Gyr) that the older they are, the higher initial masses they have.
Among them, NGC\,6838 has the lowest initial mass of $\log M_{initial} \sim$ 5.26 at the old cluster end, while at the young cluster end, Hodge\,6 has the lowest initial mass of $\log M_{initial} \sim$ 4.98.
This may give a rough initial mass threshold, if it exists, for clusters to form MPs. However, more observations of old in-situ GCs with low initial masses, e.g., Pal\,1 and Pal\,11, are still necessary before drawing any firm conclusion.

\section{Summary} \label{sec:sum}
Mass has been generally considered as one of the key parameters in deciding whether a stellar cluster will have MPs or not. In order to explore the possible mass threshold for MP formation, we studied the stellar population composition in low-mass GCs. In this work, we focus on the GC Eridanus, with a present-day mass of $1.16\times 10^{4}\,M_{\odot}$ \citep{Baumgardt2019}.

Based on its CMD, we determined the age, metallicity, distance modulus and distance for Eridanus, which are 10.0\,$\pm$\,0.5\,Gyr, -1.45\,$\pm$\,0.05\,dex, 19.83\,$\pm$\,0.04\,mag and 92.47\,$\pm$\,1.71\,kpc, respectively. These are consistent with the values derived in previous studies.

In order to investigate whether MPs exist in Eridanus, blue-UV low-resolution spectra were obtained with OSIRIS/MOS on GTC for 19 member stars to study their carbon and nitrogen abundances.
After determining their effective temperature and surface gravity photometrically, the spectral indices of CH4300 and CN4142 were computed for the final sample including nine RGB stars and two AGB stars. Then, with the help of model spectra, we derived the values of [C/Fe] and [N/Fe] for these stars.
Since most of them have evolved past the RGB bump, where extra-mixing may significantly alter their surface chemical composition, we compared the observed abundances with  stellar evolutionary models considering thermohaline instability and rotation-induced mixing. We find that besides the variations in carbon and nitrogen abundances caused by stellar evolution, the imprint of initial elemental abundance dispersion is prominently present in the observed nitrogen abundance distribution,which indicates the existence of MPs in Eridanus.

To find further investigate the critical cluster mass for MPs, we updated the age-mass distribution plane for all clusters having been studied for MPs. Although a threshold of present-day mass at $\log M \sim$ 4.65 seems to exist for clusters younger than 10\,Gyr, the condition for GCs older than 10\,Gyr is unclear. Taking the mass loss during the long-time evolution of GCs into consideration, the initial mass of clusters should be a more reasonable parameter to analyze when studying the formation of MPs. We then constructed the age-initial mass plane with clusters in the Milky Way and LMC/SMC. To avoid the influence of the uncertainty in the initial mass estimation of accreted clusters, we simplified our sample with only in-situ clusters\footnote{We note that theoretically one does not expect the formation of MPs to depend on whether or not the cluster was formed in-situ or not.}. Then, a slight trend that initial mass increases with increasing age is found for clusters (older than 2\,Gyr) with MPs, and $\log M_{initial} \sim$ 4.98 and 5.26 are the lowest initial mass at the young and old end, respectively, which might provide a reference for the critical mass for clusters to form MPs.

However, the current observations are still incomplete for the study on this topic. Other low-initial-mass in-situ clusters are still necessary to be studied before we are able to have a comprehensive understanding.
In this context, the up-coming space telescopes, e.g., the China Space Station Telescope (CSST) %\citep[CSST,][]{Zhan2011CSST,Cao2018CSST,Gong2019CSST}
, as well as the ground based next-generation large telescopes, e.g., Thirty Meter
 Telescope and Extremely Large Telescope, will provide exciting opportunities for observations and studies in this field.

\begin{acknowledgements}

Y.W. acknowledges the support by the National Natural Science Foundation of China under grant 11803048 and the Special Research Assistant Fundation Project of Chinese Academy of Sciences.
B.T. was supported by the Natural Science Foundation of Guangdong Province under grant No. 2022A1515010732 and National Natural Science Foundation of China under grant No. U1931102.
C.L. was supported by the National Natural Science Foundation of China through grants 12037090 and 11803048, Guangdong Major Project of Basic and Applied Basic Research (grant No. 2019B030302001) and the hundred-talent project of Sun Yat-sen University.
Y.W., B.T. and C.L. acknowledge the National Natural Science Foundation of China through grants 12233013, and the science research grants from the China Manned Space Project with NO. CMS-CSST-2021-A08 and CMS-CSST-2021-B03.
R. R. M. gratefully acknowledges support by the ANID BASAL project FB210003.
J.G.F-T gratefully acknowledges the grant support provided by Proyecto Fondecyt Iniciaci\'on No. 11220340, and also from ANID Concurso de Fomento a la Vinculaci\'on Internacional para Instituciones de Investigaci\'on Regionales (Modalidad corta duraci\'on) Proyecto No. FOVI210020, and from the Joint Committee ESO-Government of Chile 2021 (ORP 023/2021).
D.G. gratefully acknowledges support from the ANID BASAL project ACE210002.
D.G. also acknowledges financial support from the Direcci\'on de Investigaci\'on y Desarrollo de la Universidad de La Serena through the Programa de Incentivo a la Investigaci\'on de Acad\'emicos (PIA-DIDULS).

This work is based on observations made with the Gran Telescopio Canarias (GTC), installed in the Spanish Observatorio del Roque de los Muchachos of the Instituto de Astrofisica de Canarias, in the island of La Palma.
This work uses the data obtained from the European Space Agency (ESA) space mission Gaia.

\end{acknowledgements}

%\appendix

\bibliography{GC-Eri}{}
\bibliographystyle{aasjournal}

\end{document}